\documentclass[]{mn2e}
\usepackage[dvips]{graphicx}
\usepackage{times}
\usepackage{times,amssymb}

\newcommand{\be}{\begin{equation}}
\newcommand{\ee}{\end{equation}}
\newcommand{\bea}{\begin{eqnarray}}
\newcommand{\eea}{\end{eqnarray}}
\begin{document}
\title[Origin and Detectability of Trojan planets]{Origin and Detectability of coorbital planets from radial velocity data}
\author[C.A. Giuppone, P. Ben\'itez-Llambay and C. Beaug\'e]
{C.A. Giuppone$^{1,2}$, P. Ben\'itez-Llambay$^{1,3}$ and C. Beaug\'e$^{1,3}$ \\
$^{1}$Observatorio Astron\'omico, Universidad Nacional de C\'ordoba, Laprida 854, 
(X5000BGR) C\'ordoba, Argentina \\
$^{2}$Departamento de F\'isica, I3N, Universidade de Aveiro, Campus de Santiago, 3810-193 Aveiro, Portugal (cristian@ua.pt) \\
$^{3}$Instituto de Astronom\'ia Te\'orica y Experimental, Laprida 854, (X5000BGR) C\'ordoba, Argentina
}

\maketitle

\date{}

\begin{abstract}
We analyze the possibilities of detection of hypothetical exoplanets in coorbital motion from synthetic radial velocity (RV) signals, taking into account different types of stable planar configurations, orbital eccentricities and mass ratios. For each nominal solution corresponding to small-amplitude oscillations around the periodic solution, we generate a series of synthetic RV curves mimicking the stellar motion around the barycenter of the system. We then fit the data sets obtained assuming three possible different orbital architectures: (a) two planets in coorbital motion, (b) two planets in a 2/1 mean-motion resonance (MMR), and (c) a single planet. We compare the resulting residuals and the estimated orbital parameters.

For synthetic data sets covering only a few orbital periods, we find that the discrete radial velocity signal generated by a coorbital configuration could be easily confused with other configurations/systems, and in many cases the best orbital fit corresponds to either a single planet or two bodies in a 2/1 resonance. However, most of the incorrect identifications are associated to dynamically unstable solutions. 

We also compare the orbital parameters obtained with two different fitting strategies: a simultaneous fit of two planets and a nested multi-Keplerian model. We find that, even for data sets covering over ten orbital periods, the nested models can yield incorrect orbital configurations (sometimes close to fictitious mean-motion resonances) that are nevertheless dynamically stable and with orbital eccentricities lower than the correct nominal solutions. 

Finally, we discuss plausible mechanisms for the formation of coorbital configurations, by the interaction between two giant planets and an inner cavity in the gas disk. For equal mass planets, both Lagrangian and anti-Lagrangian configurations can be obtained from same initial condition depending on final time of integration.

\end{abstract}


\section{Introduction}

The possible existence of exoplanets in coorbital motion has fascinated planetary scientists for several years. Since the diversity of exoplanetary configurations continue to surprise us, even more than 15 years after the discovery of Peg51b, and it seems almost natural to expect Trojan planets to exist somewhere and the announcement of their discovery to be only a matter of time. Probably the first detailed analysis of hypothetical coorbital planets is due to Laughlin and Chambers (2002). They studied three types of coorbital configurations: tadpole orbits (around $L_4$ and $L_5$ equilateral points), horse-shoe configurations and ``eccentric resonances''. They proposed to distinguish between coorbital and single planet fits from RV data by observing residuals from long-term observations (more than ten orbital periods), because the coorbital configurations have large mutual interactions due to resonant motion.

For systems with more than one planet, it is well known that the existence of resonant motion may be possible evidence of a past large-scale planetary migration due to interactions with the gaseous disk. Although their importance is unquestionable, it is still intriguing why some commensurabilities are very populated (e.g., 2/1 MMR) and others that are currently empty (particularly the 1/1 MMR). 

Lagrange (1873) discovered stable solutions for three massive bodies such that at all times their relative positions are located in the vertices of an equilateral triangle of variable size ($L_4$ and $L_5$ solutions). Linear stability analyses traditionally focused on the restricted three-body problem (where one of the masses vanishes, e.g Morais 2001, Namouni et al. 1999 and references therein). Recently, Nauenberg (2002) numerically investigated the dynamical stability of general three body problem as a function of the eccentricity of the orbits and the Rouths' mass parameter. The resonant solutions were found stable when $\frac{m_1+m_2}{M_\odot} \lesssim 0.038$ and up to eccentricities $\sim 0.6$. The existence of such periodic orbits in the general three-body problem also was derived for example in Siegel \& Moser (1971) and Laughlin \& Chambers (2002). As mentioned in Laughlin \& Chambers (2002), horse-shoe orbits are stable for planetary masses up to $\sim 0.4 m_{jup}$, and despite first impression, Keplerian fit are good enough to reveal their radial velocity signature with only four orbital periods. We note also that those type of orbits named ``eccentric resonances'', are simply an effect of angular momentum conservation in a coplanar three-body problem, that predicts when the eccentricity of one planet is maximum the other eccentricity reach a minimum and they have high amplitude of oscillation of resonant angles. 

More recently Hadjidemetriou et al. (2009) studied the properties of motion close to a periodic orbit by computing the Poincar\'e map on a surface of section. They constructed the symmetric families of stable and unstable motion describing one previous unknown symmetric configuration in the general problem, the Quasi-Satellite (QS) solution.  

In a previous work (Giuppone et al. 2010), we studied the stability regions and families of periodic orbits of two planets in the vicinity of a 1/1 MMR, using numerical integrations and developing a semi-analytical method. We considered different ratios of planetary masses and orbital eccentricities, although we assumed that both planets share the same orbital plane (coplanar motion). As result we identified two separate regions of stability, symmetric and asymmetric types of motion easily described with the behavior of resonant angles $(\sigma, \Delta \varpi) = (\lambda_2 \!-\!\lambda_1 , \varpi_2-\varpi_1)$, summarizing\footnote{We recommend see the Figs. 2 and 5 from Giuppone et al. (2010) in order to easily identify these kind of coorbital motions in same plane of astrocentric parameters. We center this work on  $L_4$ and $AL_4$ configurations since, due to symmetry consideration, they give same results as $L_5$ and $AL_5$ respectively.}

\begin{itemize}
\item 
\textbf{Quasi-Satellite region:} correspond to oscillations around a fixed point located at $(\sigma,\Delta\varpi) = (0,180^{\circ})$ independent of mass ratio. Although not present for quasi-circular trajectories, they fill a considerable portion of the phase space in the case of moderate to high eccentricities.

\item 
\textbf{Lagrangian region:} Two distinct types of asymmetric periodic orbits exists in which both $\sigma$ and $\Delta\varpi$ oscillate around values different from $0$ or $180^\circ$. The first is the classical equilateral Lagrange solution associated to local maxima of the averaged Hamiltonian function. Independently of the mass ratio $m_2/m_1$ and their eccentricities, these solutions are always located at $(\sigma,\Delta\varpi) = (\pm 60^\circ,\pm 60^\circ)$. However, the size of the stable domain decreases rapidly for increasing eccentricities, being practically negligible for $e_i > 0.7$.

The second type of asymmetric solutions correspond to local minimum of the averaged Hamiltonian function. They were named Anti-Lagrangian points ($AL_4$ and $AL_5$) and, for low eccentricities, are located at $(\sigma,\Delta\varpi) = (\pm 60^\circ,\mp 120^\circ)$. Each is connected to the classical $L_i$ solution through the $\sigma$-family of periodic orbits in the averaged system (a family of solutions where the angle $\sigma$ has zero amplitude of oscillation, e.g. Michtchenko et al. 2008ab). Contrary to the classical equilateral Lagrange solution, their location in the plane $(\sigma,\Delta\varpi)$ varies with the planetary mass ratio and eccentricities. Although their stability domain also shrinks for increasing values of $e_i$, they do so at a slower rate than the classical Lagrangian solutions, and are still appreciable for eccentricities are high as $\sim 0.7$. An empirical relation exist for eccentricities below 0.6 (e.g. Giuppone et al. 2010, Hadjidemetriou \& Voyatzis 2011)
\be
e_1 \simeq \biggl( \frac{m_2}{m_1} \biggr) e_2.
\label{e1e2}
\ee
\end{itemize}

In this paper we address the question of extrasolar Trojans, both from its detectability and possible origin.
In a recent work, Anglada-Escude et al. (2010) showed that two planets in a 2/1 MMR resonance could mimic a single planet in a more eccentric orbit. We wonder whether the opposite could also occur, and if two planets in a coorbital configuration could give similar RV signals as other dynamical systems including the 2/1 resonance. 
In Section 2 we review the radial velocity signal generated by such systems and how they can be easily mistaken by the signal generated by a single planet. In the following section we select representative configurations and generate synthetics data-sets for several mass ratio. Sections 4 analyzes the dynamical stability of the different possible configurations, while the dispersion of the best-fit parameters is discussed in Section 5. A comparison between two different strategies for planetary detection (two-planet fit vs. nested Keplerian fits) is presented in Section 6. Finally, a possible formation mechanism of extrasolar coorbitals is discussed in Section 7, where we present a series of numerical simulations (both hydro and N-body) of two planet systems immersed in a gaseous disk with an inner gap. Conclusions close the paper in Section 8.

\section{Keplerian Radial Velocity equations}

First we need to understand some aspects involved in the RV signal produced by two planets in coorbital motions, assuming unperturbed Keplerian motion. Consider two planets with masses $m_1$ and $m_2$ in coplanar orbits around a star with mass $m_0=M_{\odot}$. Let $a_i$ denote the semimajor axes, $e_i$ the eccentricities, $\lambda_i$ the mean longitudes, $\varpi_i$ the longitudes of pericenter, $M_i$ the mean anomaly and $f_i$ the true anomalies. All orbital elements are assumed astrocentric and osculating.

The suitable angular variables for coorbital motion are defined by $\sigma = \lambda_2-\lambda_1$ and $\Delta\varpi = \varpi_2-\varpi_1$.  Disregarding mutual interactions between the planets, the radial velocity of the star is the sum of the individual Keplerian contributions (e.g. Beaug\'e et al. 2007). 
\bea
V_r &=& {K_1} \big(\cos{(f_1+\varpi_1)} + e_1 \cos{\varpi_1}\big) \nonumber \\
   & +& {K_2} \big(\cos{(f_2+\varpi_2)} + e_2 \cos{\varpi_2}\big) \label{vr_exact} \\
&& \mbox{where   } K_i = \frac{m_i \sin{I_i}}{m_i + m_0}\frac{n_i a_i}{\sqrt{1-e_i^2}} \nonumber  ,
\eea
$I_i$ being the orbital inclination with respect to the sky. Even though the radial velocity of a star perturbed by the two planets near the periodic orbit (i.e QS, $L_i$ and $AL_i$) is not given by a single periodic signal, we may ask under what circumstances this signal could be mimic by a single planet orbiting the star or even two planets in other configurations. 

To address this question we can use the same approach than Anglada-Escude et al. (2010) rearranging the Keplerian contributions from each planet. Using the expansions to first order in eccentricities for the true anomaly $f$ (see e.g. Murray \& Dermott, 1999)
\bea
\sin(f_i)&=&\sin(M_i)+e_i \, \sin(2M_i) \nonumber \\
\cos(f_i)&=&\cos(M_i)+e_i \cos(2M_i)-e_i ,
\eea 
and expanding
\bea 
\cos{(f_i+\varpi_i)} = \cos{(\varpi_i + M_i)} + e_i \cos{(2 M_i + \varpi_i)} - e_i \cos{\varpi_i},
\label{eq-fw}
\eea
recalling that $\lambda_i= \varpi_i + M_i$ and substituting (\ref{eq-fw}) into (\ref{vr_exact}) we finally obtain
\bea
V_r &\simeq& K_1 \biggl(\cos{\lambda_1} + e_1 \cos{(2 \lambda_1-\varpi_1)}\biggr) \nonumber \\ 
     &+& K_2 \biggl(\cos{\lambda_2} + e_2 \cos{(2 \lambda_2-\varpi_2)}\biggr)
\label{eq-4}
\eea 

\begin{figure}
 \centerline{\includegraphics*[width=18pc]{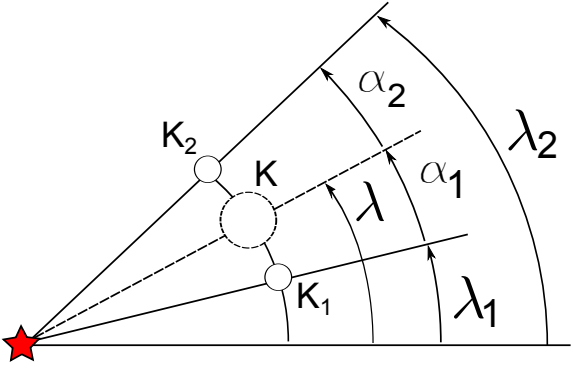}}
 \caption{Schematic configuration for coorbitals and corresponding single planet that mimic their RV signal. The position of single planet is determined by equations (\ref{eq10}) and (\ref{eq11})}.
  \label{angles}
 \end{figure}

Now, we suppose that the previous signal can be mimicked by a single planet. Imagine that its mean longitude $\lambda$ is located somewhere between the position of coorbitals. According to Figure \ref{angles} we can rewrite the mean longitudes and the longitudes of pericenters as
\bea 
\lambda_1&=&\lambda-\alpha_1 \nonumber \\
\lambda_2&=&\lambda+\alpha_2, \hspace{1 cm} \mbox{where} \hspace{1 cm}   \alpha_1+\alpha_2=\sigma \nonumber \\
\varpi_1 &=& \varpi - \beta_1 \label{eq-5} \\
\varpi_2 &=& \varpi + \beta_2 \hspace{1 cm} \mbox{where} \hspace{1 cm}  \beta_1+\beta_2=\Delta\varpi \nonumber
\eea 

By definition of periodic orbit $\dot{\sigma}=0$, $\Delta\dot{\varpi}=0$ giving constant values for $\alpha_1,\alpha_2, \beta_1 \mbox{and } \beta_2$. We can then substitute eqs. (\ref{eq-5}) into eq. (\ref{eq-4}), and regroup terms to rewrite
\be
V_r = \underbrace{(K_1 \cos{\alpha_1} +K_2 \cos{\alpha_2})}_{\equiv \mbox{K}} 
\,\Biggr(\cos{\lambda} 
     + e \, {\cos{(2 \lambda - \varpi)}}\Biggl) 
\label{eq-7}
\ee
\bea
e &\equiv& {\frac{K_1 e_1 \cos{(2 \alpha_1 - \beta_1)}+K_2 e_2 \cos{(2 \alpha_2 - \beta_2)}}{K_1 \cos{\alpha_1}+K_2 \cos{\alpha_2}}} 
\label{eq-8}
\eea
where the conditions
\bea 
K_1 \sin{\alpha_1} &=& K_2 \sin{\alpha_2} \label{eq10}\\
K_1 e_1 \sin{(2 \alpha_1 - \beta_1)} &=& K_2 e_2 \sin{(2 \alpha_2 - \beta_2)} \label{eq11}, 
\eea 
eliminate the additional periodic term and also define the position of fictitious single planet, giving
\bea 
\tan{\alpha_1} &=& \frac{K_2}{K_1} \frac{\sin{\sigma}}{(1+\frac{K_2}{K_1} \cos{\sigma})} \nonumber \\
\tan{(2 \alpha_1 - \beta_1)} &=& \frac{K_2 e_2}{K_1 e_1} \frac{\sin{(2\sigma-\Delta\varpi)}}{\biggl(1+\frac{K_2 e_2}{K_1 e_1} \cos{(2\sigma-\Delta\varpi)}\biggr)}  .
\eea
It can easily be seen that when both planets have the same mass $\alpha_1=\alpha_2$ and $\beta_1=\beta_2$, from which the fictitious single planet would have $\lambda=\frac{\lambda_1+\lambda_2}{2}$ and $\varpi=\frac{\varpi_1+\varpi_2}{2}$. In this case $e_1=e_2$ and the RV signal arising from the single planet would be
\bea 
K &=& 2 K_1 \cos{\frac{\sigma}{2}} \label{eq-simplvr0} \\
e &=& e_1 \frac{\cos{(\sigma-\frac{\Delta\varpi}{2})}}{\cos{\frac{\sigma}{2}}}\label{eq-simplvr}  .
\eea

In the opposite limit, when $m_2 \gg m_1$, both $\alpha_2,\beta_2 \rightarrow  0$ and the single planet is located near the position of the massive planet. Recall that for coorbital configuration with eccentricities up to $e_i<0.6$, the stable zero-amplitude solutions satisfy eq. (\ref{e1e2}), except for the $L_4$ configuration that always lies in the line segment $e_1=e_2$. 

\begin{figure}
 \centerline{\includegraphics*[width=23pc]{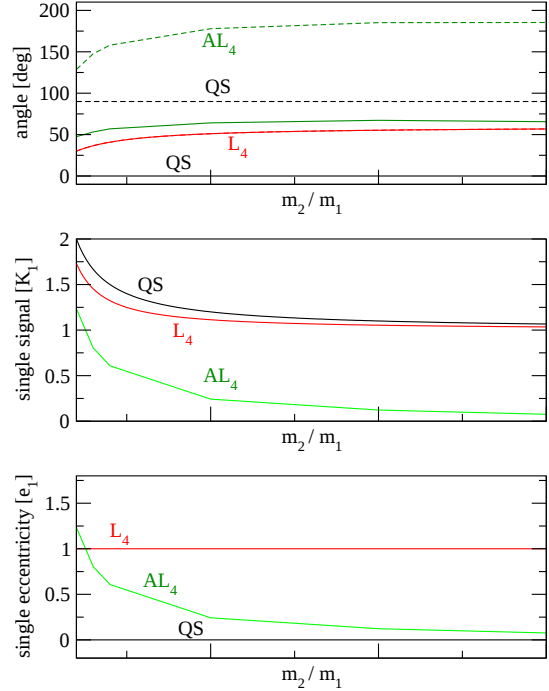}}
 \caption{Parameters for a single planet to mimicking the radial velocity signal of two coorbitals with mass ratio $m_2/m_1$. The color code is used to distinguish configurations: QS in black, $L_4$ in red and $AL_4$ in green. \textbf{Top:} Values of $\alpha_1$  are shown in continuous lines while $\beta_1$ is plotted in dashed lines. \textbf{Middle:} Associated value of $K$ in units of $K_1$. \textbf{Bottom:} Equivalent eccentricity of single planet for each configuration in units of the eccentricity of the coorbital planet with signal $K_1$.}
  \label{fig-eqs}
 \end{figure}

For other mass ratios, Figure \ref{fig-eqs} shows the equilibrium values of $\alpha_1, \beta_1$ and the corresponding signal $K$ amplitude and eccentricity $e$ (the eccentricity is shown in units of $e_1$). Note that the value of $e$ only varies when the single planet mimics coorbitals in $AL_4$. It is interesting remark that QS configuration could easily mimicked by single planet of circular orbit. 

\section{Orbital Fits and Synthetic Radial Velocity}

\begin{figure}
 \centerline{\includegraphics*[width=18pc]{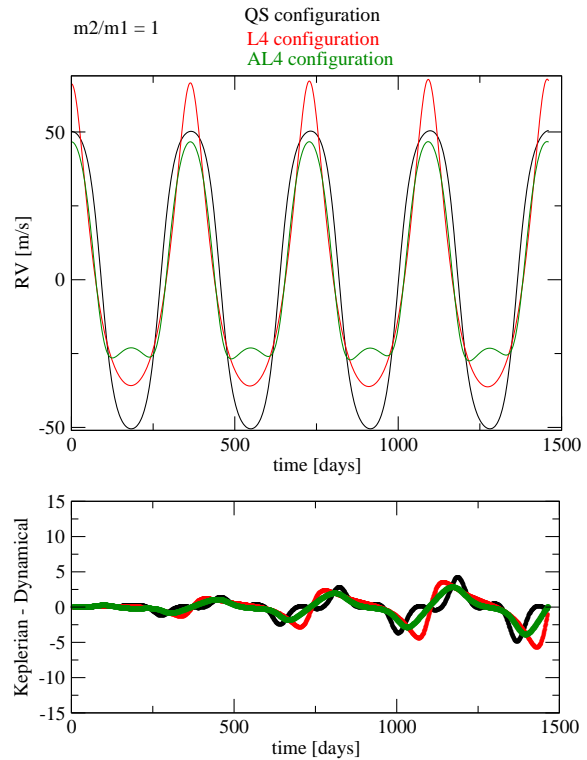}}
 \caption{\textbf{Top:} Synthetic radial velocity curves generated with an N-body integrator for three different orbital configurations, specified in the top of the graph {for the first conditions in Table \ref{tab_po}}. \textbf{Bottom:} Difference between radial velocities calculated from the N-body integration and those generated with Keplerian model. Although the difference increases with time, it remain below 4 m/s even after four orbital periods.}
  \label{VR-comp}
 \end{figure}

First we review the differences in radial velocities signals using the Keplerian approximation (\ref{vr_exact}) and N-body integrator, in order to distinguish how important are the mutual interactions and the feasibility of being detected. The top frame of Figure \ref{VR-comp} shows three synthetic curves covering four orbital periods for equally mass planets ($m_i=m_{\rm Jup}$) in QS, $L_4$ and $AL_4$ configurations. The bottom frame of the same figure shows the error that would be obtained if the RV signal was calculated assuming constant Keplerian orbits. Even after four orbital periods, the error remains below $\sim$ 4 m/s. As expected, the mutual perturbations would be even smaller if the individual mass were reduced.

To test the mimicking effect described in the previous section, we selected six nominal solutions near exact fixed points chosen from Giuppone et al. (2010), three with $m_1=m_2$ and three other with $m_1 = m_2/3$. In all cases we fixed $m_2 = 1 m_{\rm Jup}$. Initial conditions are summarized in Table \ref{tab_po}. In order to test the validity of our expressions (\ref{eq-4}) and (\ref{eq-7})-(\ref{eq-8}), constructed from a first-order expansion in the eccentricities, here we deliberately chose high eccentric orbits ($e_1 = 0.3$). If our predictions prove correct in these cases, we can be assured that they will be valid for lower eccentricities as well.  

\begin{table}
\begin{center}
\caption{Arbitrary conditions near the stable periodic solutions in the $(\sigma, \Delta \varpi)$ plane, calculated with semi-analytical method. All conditions have $a_1=a_2$.}
\label{tab_po}
\begin{tabular}{|c|c|c|c|c|c|c|c|}
\hline 
 &$m_2/m_1$ & $\sigma$ (deg) & $\Delta\varpi$ (deg) & $e_1$ & $e_2$   \\
\hline
 QS    & 1 &    0   &   180  &  0.3 & 0.3    \\
 $L_4$   & 1 &   60   &    60  &  0.3 & 0.3 \\
 $AL_4$  & 1 &   95.4   &   257.4 & 0.3 & 0.3 \\
\hline
 QS   & 3  &    0   &   180   &  0.3 & 0.1 \\
 $L_4$  & 3  &   60   &    60   &  0.3 & 0.3 \\
 $AL_4$ & 3  &   77.6 &   249.7 &  0.3 & 0.1 \\
\hline 
\end{tabular} 
\end{center}
\end{table} 

For each nominal configuration we generated a synthetic RV curve describing the stellar motion around the barycenter of the system. The curve was then represented with a discrete sampling of $N$ observation times $t_i$ distributed randomly, according to an homogeneous distribution (thus avoiding aliases in data from daily/seasonal observations, e.g. Dawson \& Fabrycky, 2010). In each data point the nominal RV value $V_r(t_i)$ was displaced to a new value ${V_r}_i = V_r(t_i) + \mathcal{N}(0,\epsilon)$ following a Gaussian distribution with constant variance $\epsilon^2$. The resulting synthetic data set was the used as input for our orbital fitting code {\it PISA} (Pikaia genetic algorithm + simulated annealing) (e.g. Beaug\'e et al. 2008). Since our data sets cover only four orbital periods, the orbital fit was obtained assuming non-interacting Keplerian orbits. 

1000 different data sets were generated for each orbital configuration (QS, $L_4$, $AL_4$) and each mass ratio $m_2/m_1$. Each synthetic data set consisted of $N=100$ data points covering a total of four orbital periods. The results shown here were obtained adopting $\epsilon=3$ m/s, a value similar to the present day errors in detections programs (including stellar jitter). The same analysis was repeated for other values of $\epsilon$, showing similar results although the dispersion around the mean values decreased/increased as function of $\epsilon$ (see e.g. Giuppone et al. 2009). 

As noticed by Laughlin \& Chambers (2002) and Gozdziewski \& Konacki( 2006), two planets in a coorbital configuration produce only one peak in a Fourier spectrum, meaning that it could easily be confused with a single planet (obviously more massive than the individual components). On the other hand,  Anglada-Escude et al. (2011) found that two planets in circular orbits near a 2/1 MMR could also be falsely detected as a single planet in an eccentric orbit. So, our question here is the following: is it possible for two planets in different coorbital configurations (QS, $L_4$ or $AL_4$) to be falsely identified as a single planet or as two bodies in a 2/1 resonance? 

\begin{figure}
\centerline{\includegraphics*[width=18pc]{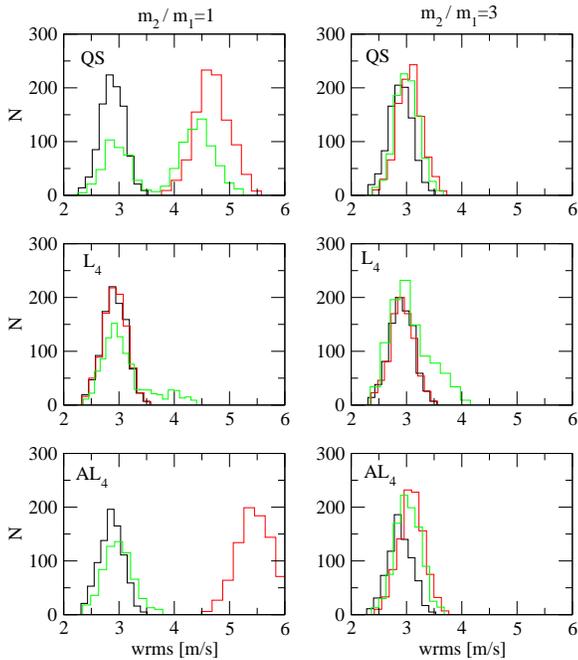}}
\caption{Histograms of residuals adopting different orbital configurations to fit a series of synthetic RV signals. The nominal solution is shown in the top left-hand side of each frame. Each colored histogram correspond to a different assumed orbital configuration: two planets in coorbital motion (black), a single planet (red) and two planets in a 2/1 MMR (green).}
 \label{rms-stat}
\end{figure}

To test this idea, each of the synthetic data sets was fitted assuming: (i) a single planet, (ii) two planets in coorbital configuration (without specifying the type) and (iii) two planets in the vicinity of a 2/1 MMR. The resulting values of {\it wrms} (weighted rms) for the 1000 data sets associated to each nominal solution are displayed as histograms in Figure \ref{rms-stat}. Each frame corresponds to a given nominal solution and mass ratio, while the color code for the histograms corresponds to the fitted orbital configurations. For example, in the top left-hand plot we constructed 1000 synthetic data sets from a nominal quasi-satellite configuration of two planets with equal masses. The data sets where then fitted assuming the three different possible 
configurations,  the resulting values of {\it wrms} plotted. The colored histograms show the distribution of the residuals of each assumed configuration. As we can see, the results corresponding to QS solutions show to smallest distribution of {\it wrms} (both in mean and dispersion), very similar to the adopted value of $\epsilon$. Recall that ${wrms} \sim \epsilon$ is usually employed as evidence that not only the fit is satisfactory but also that the assumed planetary model is correct.

Although for nominal QS solutions and $m_1=m_2$ the coorbital configuration gives the best fits, this is not always the case. For systems with a more massive outer body ($m_2/m_1=3$), and for any nominal coorbital motion, we find that all three assumed orbital solutions given practically the same error distribution (right-hand side plots). In other words, it seems that for $m_2 > m_1$ the difference in short-term RV signals of one planet or two bodies in either a 1/1 or 2/1 MMR are virtually indistinguishable.  For equal-mass planets, the picture is similar, specially if the nominal configuration corresponds to two planets in $L_4$. However, two bodies in other coorbital configurations (i.e. QS or $AL_4$) appear easier to identify.

\begin{figure}
\centerline{\includegraphics*[width=18pc]{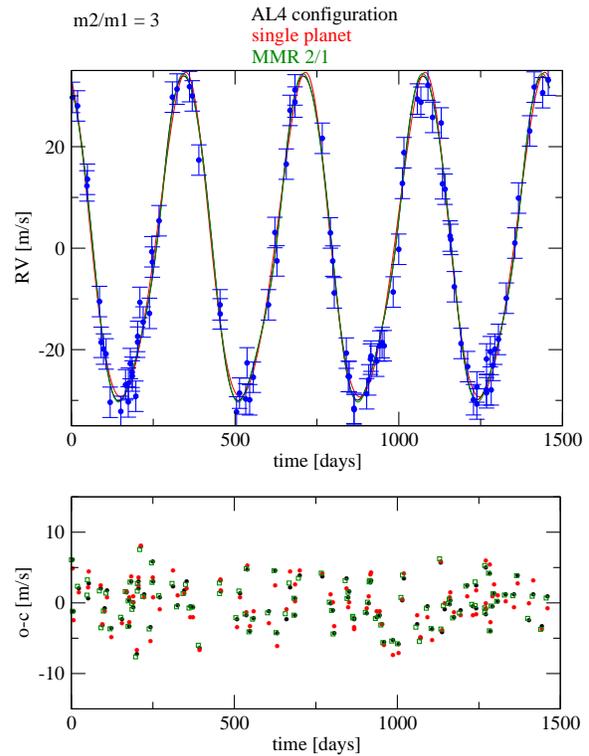}}
\caption{Comparison of synthetic curves (top frame) and residuals (bottom frame) from a nominal $AL_4$ solution and fitting the synthetic data set with three planetary models: coorbitals (black), single planet (red) and 2/1 MR (green). The best fit values of the masses and orbital elements are shown in Table \ref{tab_al4}.}
 \label{fig-al4}
\end{figure}

Figure \ref{fig-al4} shows an example. The top frame shows a synthetic data set constructed from a nominal solution of two planets in a stable $AL_4$ configuration. The three different fits are superimposed in different colors and show practically no difference. The corresponding distribution of residuals are shown in the bottom frame, also displaying no significant difference. The best fit planetary masses and orbital elements of each orbital configuration are shown in Table 2. 

\begin{table}
\begin{center}
\small
\caption{Solutions from synthetic data sets generated from configuration $AL_4$ in Table 1.}
\begin{tabular}{|c|c|c|c|c|c|c|c|}
\hline 
  fit       &  $m_1$ & $m_2$ & $a_1$  & $a_2$ & $e_1$ & $e_2$ & wrms   \\
 model       &  $m_{jup}$ & $m_{jup}$ & AU & AU &  &  & m/s   \\
\hline
 $AL_4$          &   0.37   &   0.9  &  1 & 1 & 0.31  & 0.127 & 3.03 \\
 1 Planet   &   1.12   &   --   & 1 & -- & 0.107 & -- & 3.26\\
 2/1 MMR         &   0.18   &   1.14 & 0.63 & 1 & 0 & 0.18 & 3.05\\
\hline 
\end{tabular} 
\end{center}
\label{tab_al4}
\end{table} 

\section{Stability analysis of the fits}

Although a given coorbital pair may be falsely identified as another orbital configuration, we still do not know whether all possible alternatives are equally dynamically stable. To analyze this point, we numerical integrated each of the best fits for a time span covering $10^4$ orbital periods, and calculated the amplitude of oscillation of the resonant angle $\sigma$ as well as the difference in longitudes of pericenter $\Delta \varpi = \varpi_2 - \varpi_1$. In the case of coorbital motion the resonant critical angle is basically the synodic angle $\sigma = \lambda_2 - \lambda_1$, while for the 2/1 MMR we have chosen the so-called principal resonant angle $\sigma = 2\lambda_2 - \lambda_1 - \varpi_1$.  

The Table \ref{tab-dynamics} presents a statistical analysis of the dynamical evolution of all the best fits. We denote the dynamical behavior as ``resonant'' if the system is stable during the full integration span and both $\sigma$ and $\Delta \varpi$ show small-to-moderate amplitudes of oscillation ($< 50^\circ$) around the fixed point. We call {\it librators} those configurations where the resonant angle librates but the secular angle circulates. ``Non-resonant'' configurations are those for which both angles circulate. Finally, ``unstable'' configurations are those that led to collisions or escapes of one of the bodies within the integration span. 

 The left-hand side of Table \ref{tab-dynamics} describes the dynamics of the coorbital fits. For both mass ratios, the QS solutions appear the most robust. For equal mass planets, all the fits correctly yield stable resonant quasi-satellite orbits, fully consistent with the nominal solution. For $m_2/m_1=3$, however, $13 \%$ of the synthetic data sets led to unstable coorbital solutions. The reliability of the fits decreases when the adopted nominal solution is $L_4$ and even worse when it is $AL_4$. In the latter case, more than a third of the data sets led to unstable solutions. 

The right-hand side of the Table 3 now shows results for those fits that incorrectly identified the signal as that generated by two planets in the vicinity of the 2/1 MMR. For equal mass planets, the best fits are highly unstable for both QS and $L_4$ nominal configurations, while the $AL_4$ data give a 63 \% of stable configurations but with both resonant angles circulating. When we increase the mass ratio, the compatible 2/1 MMR solutions are sightly more stable although again correspond mainly to non-resonant configurations. Again, the QS and $L_4$ configurations are the most unfavorable for being labeled as 2/1 MMR solutions, although for asymmetric equilateral Lagrange solutions, a small fraction  ($ \sim 8 \%$) of 2/1 MMR solutions show stable librators. 

\begin{table*}
\begin{center}
\caption{Percentage of different dynamical outcomes of the best fits for different mass ratios $m_2/m_1$. The value of $m_2$ was fixed at $1 m_{\rm Jup}$. See text for a detailed description.}
\label{tab-dynamics}
\begin{tabular}{|c|c|c|c|c|c|c|c|c|c|c|c|}
\hline 
\footnotesize
     & ${m_2}/{m_1}$  & &  \multicolumn{2}{c||}{Coorbital fit} & & & & \multicolumn{4}{c|}{2/1 MMR fit}  \\
\hline 
         &  & &  Resonant & Unstable  & & &  & Resonant & Librators & Non-resonant  & Unstable \\
         & &  & \% &  \% & & & & \%  &  \%  &  \% & \% \\
\hline 
 QS     & 1 &  &   100  &    0   & & & &   0 &   0 &    2 &  98    \\
 $L_4$    & 1 &  &     98  &    2   & & & &   0 &   7 &    3 &  92    \\
 $AL_4$  & 1 &  &     84  &  13   & & & &   0 &   0 &  63 &  37    \\
\hline 
 QS     & 3 &  &     83  &  13   & & & &   0 &   3 &  16 &  81    \\
 $L_4$    & 3 &  &     81  &  18   & & & &   0 &   6 &    7 &  87    \\
 $AL_4$  & 3 &  &     62  &  38   & & & &   0 &   8 &  31 &  61    \\
\hline 
\end{tabular} 
\end{center}
\end{table*}

\section{Dispersion of the best-fit parameters}

Although the best fits assuming two planets in a 2/1 MMR have proved to be largely unstable, and thus difficult to confuse with coorbital planets, we have no clear indication of the cause of the instability. In this section we will analyze the planetary masses and eccentricities obtained form the diverse best fits, and compare them with the nominal solutions. 

\begin{figure}
\centerline{\includegraphics*[width=18pc]{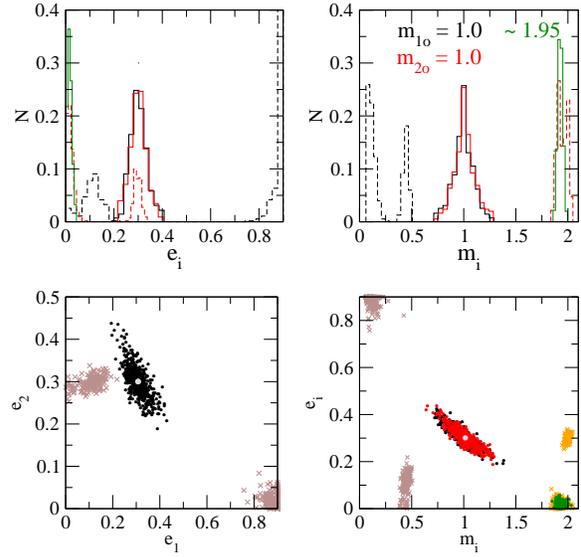}}
\caption{Dispersion of best-fit solutions from synthetic data sets for two planets with $m_1=m_2=m_{\rm Jup}$ in a QS solution. {\bf Top:} Distribution of the best-fit eccentricities (left) and planetary masses (right). The single-planet fit is shown in a continuous green line. For the two-planet fits, the results for planet $1$ are shown in black, while those corresponding to planet $2$ are shown in red. In these cases, continuous lines are used for the coorbital configuration, while dashed lines correspond to a 2/1 MMR fit. {\bf Bottom left:} Scatter plot showing the dispersion of best-fit eccentricities. Results for the coorbital fit are shown in black, while those corresponding to a 2/1 MM fit are shown in gray. {\bf Bottom right:} Scatter plot with the dispersion of eccentricity as function of planetary mass. Black and red dots correspond to both planets of a coorbital solution, orange and gray correspond to the planets of a 2/1 MMR and green points are the single planet solutions. Gray circles were drawn to identify nominal configurations.}
\label{histos_mm11qs}
\end{figure}

\begin{figure}
\centerline{\includegraphics*[width=18pc]{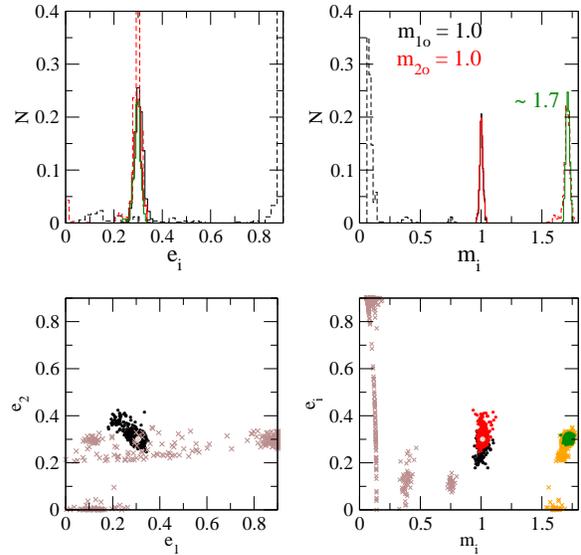}}
\caption{Same as previous figure, but for two planets with $m_1=m_2$ in an $L_4$ configuration.}
\label{histos_mm11L4}
\end{figure}

\begin{figure}
\centerline{\includegraphics*[width=18pc]{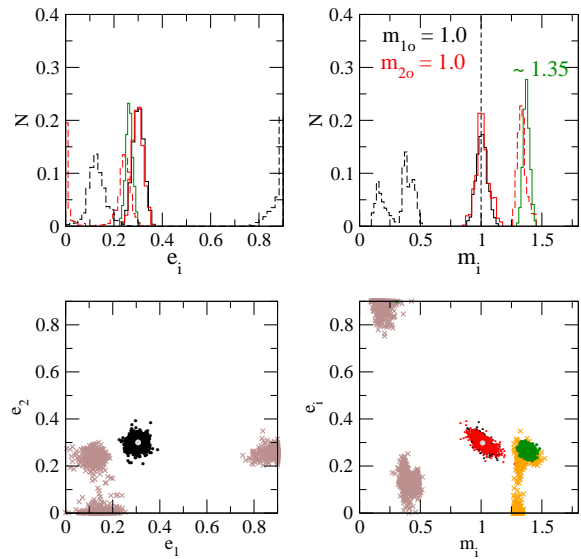}}
\caption{Histograms of fitted synthetic data sets for two equally mass planet in $AL_4$ configuration.}
\label{histos_mm11al4}
\end{figure}

Results are presented in Figures \ref{histos_mm11qs}-\ref{histos_mm11al4} for equal mass planets ($m_1=m_2=m_{\rm Jup}$). Each corresponds to a different nominal coorbital solution (QS, $L_4$ and $AL_4$, respectively), and is divided into four frames. The top graphs show histograms with the distribution of the eccentricities (left) and deduced planetary masses (right). Masses are in units of $m_{\rm Jup}$. Different colors and line types correspond to different configuration or bodies (see caption for details). The two bottom plots show the dispersion of the eccentricities (left) and relation between the mass and eccentricities (right). Once again, different colors are used for different configurations/bodies. 

For a nominal QS configuration (Figure \ref{histos_mm11qs}), the best-fit parameters assuming a coorbital solution show a considerable dispersion: approximately $0.04$ in eccentricities and $0.1 m_{\rm Jup}$ in masses. However, there appears to be no appreciable systematic error and both distributions seem symmetric with respect to the nominal values. The single-planet fits show a very steep distribution around $e \simeq 0$ and $m \simeq 2$. To understand this result, we can analyze the expected RV signal. For a QS configuration we have $\sigma = 0$ and $\Delta\varpi = 180^{\circ}$. Moreover, from equation (\ref{e1e2}) we have that equal mass planets imply $e_1=e_2$, from which we expect $K_1=K_2$. Consequently, equation (\ref{eq-7}) leads to 
\be 
V_r = 2 K_1 \cos{\lambda}. 
\label{eq:ap-qs}
\ee
This tells us that two planets in QS orbit produce the same signal as one planet with semi-amplitude $K=2 K_1$ and eccentricity $e=0$. The dispersion of the best fits show an excellent agreement with this prediction, yielding a planet with $m=m_1+m_2$ in a quasi-circular orbit. Surprisingly, the dispersion of the different fits is smaller than obtained assuming (correct) QS orbits.

The distribution of the 2/1 MMR fits, shown in the histograms with dashed lines, show a strong bimodal shape. Approximately half of the best fits give masses in the vicinity of $(m_1,m_2) \sim (2,0.5)$ and eccentricities around $(e_1,e_2) \sim (0,0.3)$. The other half of the solutions clutter around $(m_1,m_2) \sim (2,0.1)$ and
$(e_1,e_2) \sim (0.1,0.8)$. This indicates the existence of two local minima in the residual function with similar values of the {\it wrms} and similar extension in the parameter space. Small differences in the synthetic data sets would then highlight one or the other, thereby yielding either two planets in low-to-moderate eccentricities or a solution in which one of them is almost parabolic. However, as shown in Table \ref{tab-dynamics}, practically all solutions are unstable. 

Results for a nominal $L_4$ configuration are shown in Figure \ref{histos_mm11L4}. Contrary to the previous case, the coorbital fits now show a very narrow dispersion around the correct solution, implying that the RV signal of two planets in an equilateral configuration is more robust than two bodies in QS. Applying equation (\ref{eq-7}) to this case yields $\alpha_1= \alpha_2=30^\circ$, and the same RV signal could also be associated to a single planet with
\be 
V_r = {\sqrt{3}} K_1 \bigr[ \cos{\lambda} + e \cos{(2\lambda-\varpi)} \bigl]
\label{eq:ap-l4}
\ee 
where $e=e_1=e_2$ is the value of the eccentricity of the original bodies. The distribution of the best single-planet fits follow this prediction and show a very sharp peak at $(m,e)=(\sim 1.7,0.3)$. 

The 2/1 MMR best fits always give two planets with a large mass ratio. The largest body has mass and orbital elements similar to the single planet fit, while its companion has very small mass and very high eccentricity (K $\rightarrow \sigma$ and e $\rightarrow$ 1). Thus, this second planet does little more than attempt to resolve the residuals of the single planet solution. It is therefore no surprise that practically all resonant configurations are highly unstable. 

Finally, Figure \ref{histos_mm11al4} shows the results obtained from a nominal $AL_4$ configuration. Initial conditions show a small amplitude oscillation around the equilibrium solution $(\sigma,\Delta\varpi) = (95.4 ^\circ , 257.46^\circ)$ which, substituting into equation (\ref{eq-7}), leads to an equivalent single planet solution with 
\be 
V_r = 1.34 K_1 \bigr[ \cos{\lambda} + 1.24 \, e_1 \cos{(2\lambda - \varpi)} \bigl] ,
\label{eq:ap-Al4}
\ee
which results in the strong peak in the mass and eccentricity histograms. However, all one-planet solutions have much larger residuals than the coorbital solution (see Figure \ref{rms-stat}). Once again, the solutions assuming a 2/1 MMR generate a bimodal distribution, with part of the solutions ($\sim 33\%$) with high eccentricities and consequently unstable. However, contrary to the previous cases, almost two thirds of the fits lead to configurations that are dynamically stable, even if the resonant angles are in circulation. These correspond to  $m_1 \sim 0.4$ and $e_1 < 0.2$. 

The same procedure was also followed for synthetic data sets with $m_2/m_1=3$ and $e_2=e_1 /3$. Results showed very similar traits as those presented for equal-mass planets, except for a larger dispersion in the parameters. This larger dispersion diminishes the proportion of stable coorbital fits (see Table \ref{tab-dynamics}) but also allows for a significant number of stable fits assuming two planets in a  2/1 MMR. In the case of a nominal $AL_4$ configuration, almost one third of the synthetic data sets gave stable 2/1 configurations, although only a fraction of these corresponded to libration.

\section{Nested vs. two-planet fits}\label{long}

Apart from possible misidentification of coorbital planets as other configurations, the 1:1 MMR also appears sensitive to the strategy adopted for the fitting algorithm. There are two commonly used procedures to fit a two-planet solution into a given RV data set. One possibility, the {\it simultaneous two-planet fit}, assumes the existence of two masses from the beginning, and attempts to fit the data set with a model with two periodic signals. The second approach, sometimes referred to as the {\it nested model} first attempts to fit one planet into the data. If the residuals are too large or if its Fourier spectrum shows a significant periodicity, then a second planet is fitted into the reduced data. This typically is adopted if the largest amplitude of the spectrum is larger than a given false alarm probability (FAP). 

In a perfect world, both procedures should give the same results, or at least very similar to each other. This appears to be the case for non-resonant planets or even for planets in the 2/1 MMR (see the analysis done by  Beaug\'e et al 2008 for the HD82943 system). However, as we will show below, they can lead to completely different solutions in the case of coorbital planets. 

Using the same type of nominal configurations shown before, we generated new synthetic RV data sets. We incorporated two differences: first, the orbital period of the planets was raised to 400 days. Second, the data covered 12 years (10 complete periods) with $N=200$ randomly distributed observations assuming $\epsilon$=5 m/s. We repeated the same test fits as in the previous section with similar results.

\begin{figure}
\centerline{\includegraphics*[width=19pc]{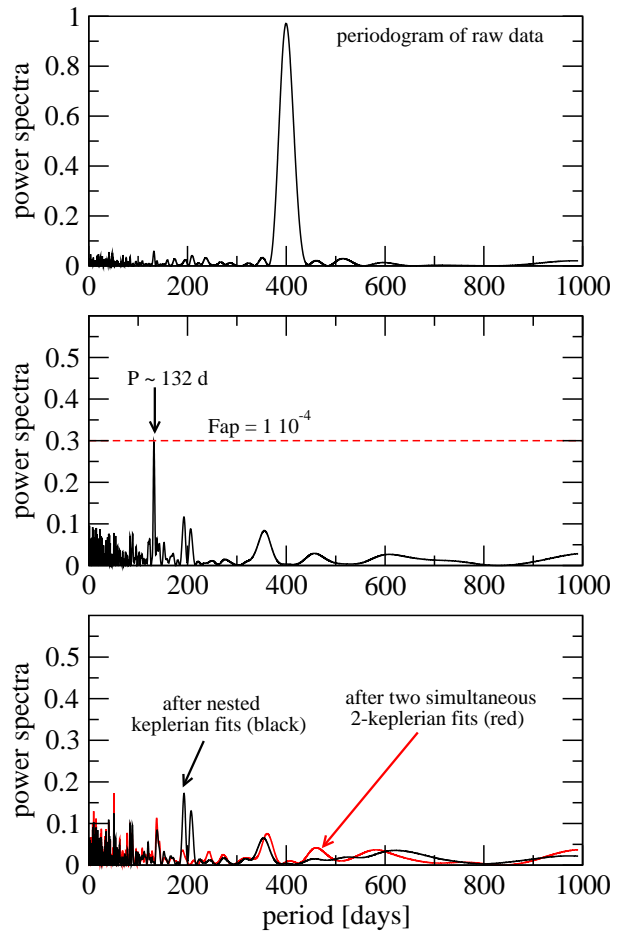}}
\caption{{\bf Top:} Power spectra of a synthetic data of two coorbital planets with orbital period of $400$ days. {\bf Middle:} Power spectra of the residuals after a single-planet fit. Horizontal dashed lines marks the $99.99 \%$ confidence level. {\rm Bottom:} Power spectra obtained from the residuals of two different two-planet fits. Red correspond to a simultaneous  fit of two massive bodies, while the black curve is the results of a nested two-planet model. }
 \label{fig-periodograma}
\end{figure}

We now use the same data to compare both fitting strategies. Results are shown in Figure \ref{fig-periodograma}. All the periodograms were calculated using DCDFT (Ferraz-Mello 1981) which allows a treatment of unequally spaced data. In the top and middle frames the power spectra are normalized so that the total area is unity. The top graph shows the periodogram of the original data, where the main $400$ day signal is clearly visible.  The middle frame shows the power spectra of the residuals after a single planet fit. The dashed horizontal line corresponds to a false Alarm Probability of ${\rm FAP} = 10^{-4}$, and was estimated using the Quast algorithm (Ferraz-Mello \& Quast 1987). We note the existence of peak corresponding to a period near $P={1\over 3} P_1$ which appears statistically relevant. After performing a second planet fit the \textit{wrms} was reduced from $6.6$ m/s to $5.6$ m/s. 

Finally, the lower frame shows periodograms of the residuals after both two-planet fits. Black curves show the results employing a nested model, while the red curve correspond to a simultaneous two-planet model. The power spectra are quite similar and no additional periodicity is observed. The resulting masses and planetary parameters of both models are given in Table \ref{tab-fits}. Although both results are very different, not only are the \textit{wrms} very similar, but the incorrect system derived from the nested model is also dynamically stable. In fact, the orbital eccentricities derived from the nested model are actually significantly lower than the nominal values. Finally, we found no appreciable difference by using N-body fits for this data set.

\begin{table}
\caption{Best multi-Keplerian orbital fits for the data generated by two equal mass planets in a QS coorbital configuration. The time of passage through the pericenter $\tau$ is given with respect to the time of the first data point. All orbital elements as astrocentric and osculating. }
\begin{center}
\begin{tabular}{|l|r|r|r|r|}
 \hline 
 \hline 
 & \multicolumn{2}{c}{Simultaneous 2-Planet Fit} & \multicolumn{2}{c}{Nested 2-Planet Fit} \\
\hline 
Parameter & Planet 1 & Planet 2 & Planet 1 & Planet 2 \\
\hline
$K$ [m/s]       & 28.59   &  31.00  &  54.19   &   5.26  \\
$P$ [d]         & 396.9  &  402.2 &  399.62  &  132.25  \\
$e$             & 0.3421   &  0.2696    &  0.0194    &    0.1835  \\
 \hline 
$m$ [$m_{Jup}$] &  1.8757   &  1.8469   &  1.96    &     0.13  \\
$a$ [AU]        &  0.7461   &  1.1779   &  1.06    &    0.508  \\
\hline
$V_{0}$ [m/s]  & \multicolumn{2}{c}{-0.051} & \multicolumn{2}{c}{0.05} \\
wrms [m/s]      & \multicolumn{2}{c}{ 5.004} & \multicolumn{2}{c}{5.62} \\
$\sqrt{\chi_\nu^2}$ & \multicolumn{2}{c}{1.027} & \multicolumn{2}{c}{1.154} \\
\hline
\hline
\end{tabular} 
\end{center}
\label{tab-fits}
\end{table} 

\begin{figure}
\centerline{\includegraphics*[width=20pc]{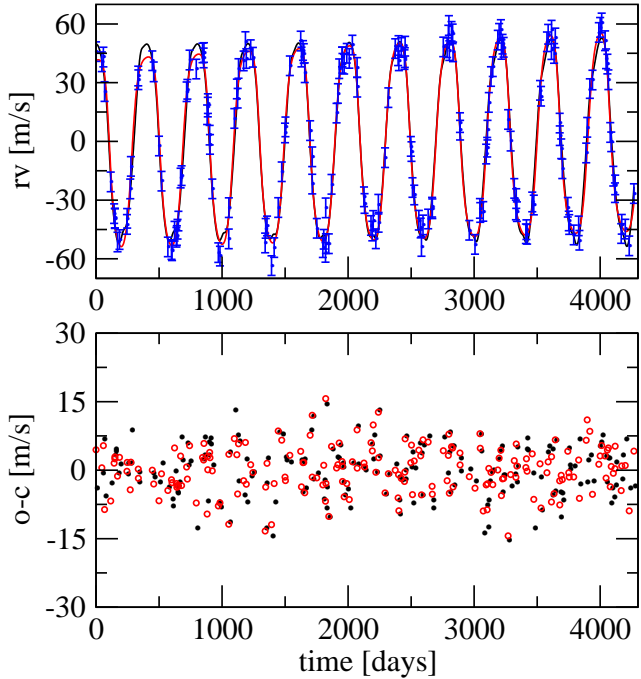}}
\caption{{\bf Top:} Synthetic data set of observations (marked as blue dots) and two curves generated with solutions obtained with coorbital simultaneous fit (red) and two nested single fits (black). {\bf Bottom:} Residuals from both fits showing no significant differences.}
 \label{fig-qslong}
\end{figure}

The top frame of Figure \ref{fig-qslong} shows, in blue, the data synthetic points of the fictitious QS configuration. The black curve presents the expected RV signal from a nested two-planet fit, while the red curve shows corresponds to a simultaneous fit of both planets. The bottom plot gives the resulting ${\rm o-c}$ showing practically no difference between both fits. 

\section{Origin of exotrojans}\label{origin}

In the first part of this paper we have analyzed whether a coorbital configuration could give RV signals similar to other dynamical systems, which may explain why so far no coorbitals have been discovered. In this second part we focus on a possible formation mechanism. 

There is a vast literature trying to explain the formation of Trojan planets. Several different mechanisms have been proposed, including accretion from protoplanetary disc (Laughlin \& Chambers 2002) pull-down capture into the 1/1 resonance, direct collisional emplacement, and in situ accretion (Chiang \& Lithwick 2005), or convergent migration of multiple protoplanets (Thommes 2005, Cresswell \& Nelson 2006). 

Kortenkamp (2005) studied the gravitational scattering of planetesimals by a protoplanet, revealing that a significant fraction of scattered planetesimals  (between 5 to 20\%) can become trapped as QS in heliocentric 1/1 coorbital resonance with the protoplanet. They included a solar nebula gas drag and considered planetesimals with diameters ranging from $\sim$1 to $\sim$1000 km. The initial protoplanet eccentricities were chosen from $e = 0$ to $0.15$ and protoplanet masses range from 300 Earth-masses ($m_\oplus$) down to 0.1 $m_{\oplus}$.

A different possibility was analyzed by Cresswell \& Nelson (2006), where the authors used a hydrodynamical code to simulate the evolution of systems of up to 10 planets with masses between  $2 m_\oplus$ and $20 m_\oplus$. After brief period of chaotic interaction characterized by scattering, orbital exchanges and collisions, some cases led to coorbital planets occupying either horse-shoe or tadpole orbits. If the initial separation between bodies was taken $\Delta \sim 5 R_{\rm Hill}$ such configurations occurred in $20\%$ of the runs, while the probability increased to $80 \%$ if the initial separation was reduced to $\Delta = 4 R_{\rm Hill}$. They also found that planets sometimes evolved from horse-shoe to tadpole configurations, although some coorbital configurations were shot-lived, ultimately leading to disruption and escapes of one of its members.

Beaug\'e et al. (2007) studied the in-situ formation of terrestrial-like Trojan planets starting from a swarm of planetesimals initially located around the Lagrangian point $L_4$ of a massive planet. They analyzed different masses, initial conditions and both gas-free and gas-rich scenarios. The swarm of proto-trojans were initially located within the $L_4$ tadpole region with total masses ranging from $1$ to $3 m_\oplus$. The gas interaction was modeled using linear and non-linear gas drags in a N-body code, adjusting the parameters to reproduce preliminary tests done with the FARGO hydrodynamical code. Their main results showed that the accretional process within the coorbital region is not very efficient, and the mass of the final Trojan planet never seems to exceed 0.6 $m_\oplus$. 

Hadjidemetriou \& Voyatzis (2011) studied the evolution of a QS system under the additional effects of a gas drag. They found that a QS planetary system with initially large eccentricities can migrate along the family of periodic orbits and be finally trapped in a satellite-type orbit. Thus the authors provided a mechanism for the generation of satellite systems starting from a planetary configuration. These results agreed with those obtained numerically by Kortenkamp (2005).
 
Morbidelli et al. (2008) complete this brief picture, analyzing the interactions between several planetary cores and a ``density jump'' in a gaseous disk. Depending on the planetary masses and initial conditions, they found cases of resonance trapping, scattering and even the formation temporary binaries. 

\begin{figure}
\centerline{\includegraphics*[width=21pc]{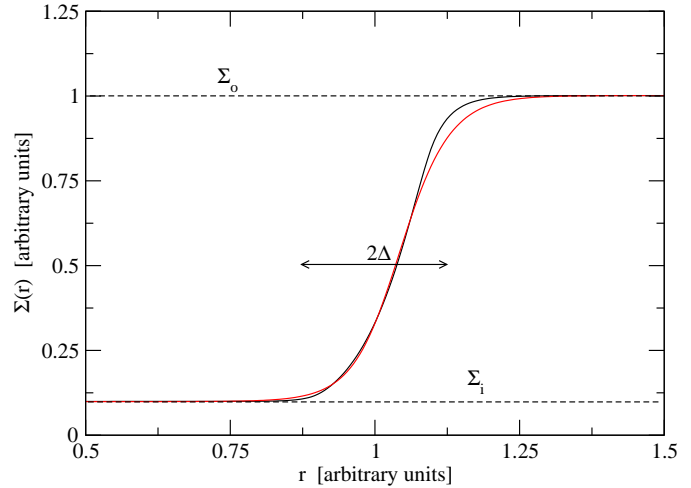}}
\caption{Surface density of the gas disk in the vicinity of a density jump or inner cavity. $\Sigma_o$ and $\Sigma_i$ are the values outside and inside the jump, also characterized by a half-width $\Delta$. The black curve shows the azimuthal average $\Sigma(r)$ obtained from a hydrodynamical simulation (FARGO) while the red curve is the analytical approximation using equation (\ref{eq-sigma}).}
 \label{cavidad}
\end{figure}

In this paper we follow the same scenario of Morbidelli et al. (2008) to analyze whether coorbital systems may also be formed through the interaction of two planets with a density jump in protoplanetary disk. We thus consider two planets initially far from mean-motion commensurabilities that migrate inwards, ultimately becoming locked in resonance. We assume that the disk possesses an inner cavity which acts as a planetary trap, halting the migration of the massive bodies. For simplicity, we consider a constant surface density $\Sigma_o$ outside the cavity, and a constant value $\Sigma_i$ inside. The density jump is set at a radial distance $r=1$ (arbitrary units) from the star. Thus, we characterize the density jump in the disk by two main parameters: the density ratio $F=\Sigma_o/\Sigma_i$ and the width $\Delta$ of the cavity edge (see Figure \ref{cavidad}). 

Our main simulations were performed using the FARGO-2D public hydro-code, whose basic algorithm may be found in Masset (2000). This code solves the Navier-Stokes equations for a Keplerian disk subject to the gravity of the central object and that of embedded planets. The disk is assumed isothermal and not self-gravitating.

\subsection{Preliminary N-body simulations}

Before undertaking the main hydro-simulations, we wish to analyze what disk properties ($\Sigma_o$, $F$, $\Delta$, etc.) and planetary parameters (masses, initial separation, etc.) are compatible with the formation of coorbital solutions. Since such a broad sweeping of the parameter space is extremely time-consuming for a hydro-code, we employed an N-body code for a series of preliminary runs. The interaction between the planets and the disk was approximated by a Type I migration following the semi-analytical model of Tanaka et al. (2002) and Tanaka and Ward (2004). In the case of a smooth density profile, explicit expressions for the forces acting on the planetary masses can be found in Ogihara et al. (2010). 

Since we will be working with planetary masses smaller than one Saturn-mass, Type I migration is expected to be a fair description of the tidal interactions with the disk. However, we stress that these N-body simulations are not expected to be accurate, but are used solely as guidelines for the full hydro-simulations introduced in the following sections. 

The inner cavity in the disk may be approximated with an hyperbolic tangent, such that
\be 
\ln{\Sigma(r)} = C_a \tanh{x} + C_b \hspace*{0.5cm} {\rm with} \hspace*{0.5cm} x = \frac{r -1}{\Delta}
\label{eq-sigma}
\ee
and where the coefficients are given by
\be
C_a = \frac{1}{2}(\ln{\Sigma_o} - \ln{\Sigma_i}) \hspace*{0.5cm} ; \hspace*{0.5cm}
C_b = \frac{1}{2}(\ln{\Sigma_o} + \ln{\Sigma_i}) .
\label{eq-cacb}
\ee
Figure \ref{cavidad} shows two representations of $\Sigma(r)$ with a test cavity (edge at $r=1$). The black curve was generated with FARGO (see Ben\'itez-Llambay et al. 2011 for details), while the red curve was obtained applying equation (\ref{eq-sigma}). Both appear very similar, although the analytical function has a smoother trend near the edges of the cavity. 

Having an explicit functional form for the density profile, we can modify Tanaka's equations for the corotational torque to reproduce the contribution generated by a density jump in the disk. This was done following the  calculations deduced in Masset et al. (2006a), and the resulting expressions were incorporated into our N-body code. With this tool we were then able to reproduce the dynamical behavior of massive planets near the density jump without needing to artificially halt the planet at a given orbital distance from the star.

Ben\'itez-Llambay et al. (2011) showed that the real population of close-in exoplanets is consistent with a disk inner cavity located at $r_0 \sim 0.01$ AU. However, since our N-body simulations will be used primarily as a stepping stone into hydro-simulations, it is preferable to scale the initial conditions to $r_0=1$. The Appendix shows the scale transformations necessary for the gas density and time units to ensure the same dynamics.
For example, assuming a typical surface density for the disk of $\Sigma \sim 5800$ gr/cm$^{2}$ at $r=1$ AU, 
the corresponding value at any arbitrary spatial scaling parameter $r_0$ gives $\Sigma=6.4 \times 10^{-4} M_{\odot} r_0^{-2}$, where $r_0$ is given in AU.

\begin{figure}
\centerline{\includegraphics*[width=21pc]{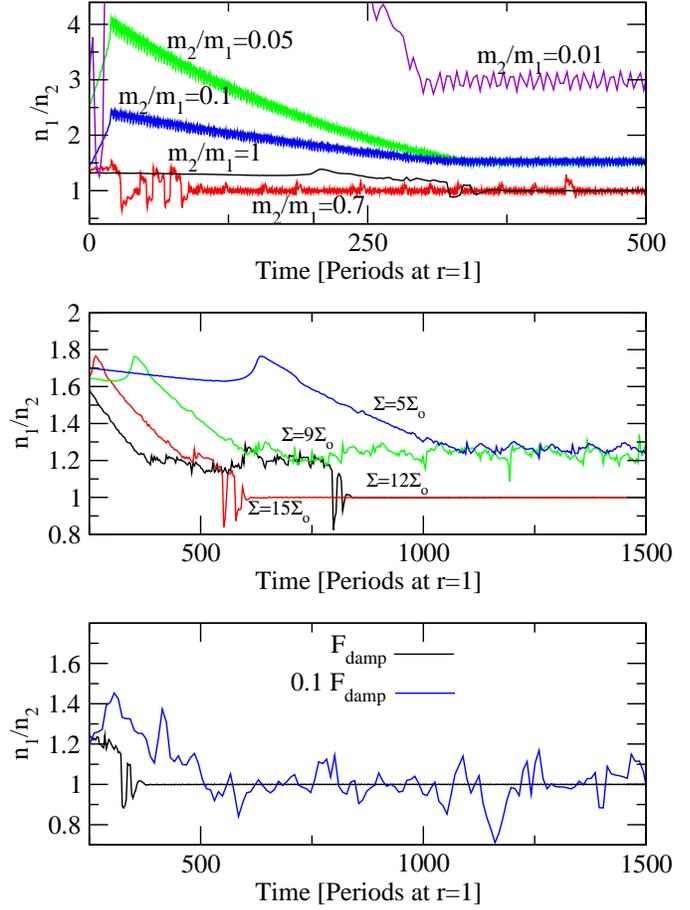}}
\caption{N-body simulations of two planets encountering an inner cavity in the disk located at $r=r_0=1$ (arbitrary units). \textbf{Top:} Using constant value for density ($\Sigma_o=15 \times \Sigma_0$ and different pairs of planets with mass ratio from $10^{-2}$ to 1. The coorbital configuration is achieved if the mass ratio is $0.7 < m_2/m_1 < 1.4$, this range becoming wider for increasing values of $\Sigma_o$. \textbf{Middle Frame.} Two Saturn-mass planets in a disk with different $\Sigma_o$. For values larger than $12\Sigma_0$ final coorbital configurations appear with high probability. \textbf{Bottom Frame.} Two simulations with $m_1=m_2=m_{\rm Sat}$ and  $\Sigma_o=15\Sigma_0$, but considering different magnitudes for the eccentricity damping. Meanwhile planet 1 was fixed at 1.5 AU, we change the position of planet 2 from 1.9 to 3.2 AU in order to represent better the results in the figure.}
 \label{fig1_p}
\end{figure}

Figure \ref{fig1_p} shows the results of several simulations using our modified N-body code with a density jump centered at $r=r_0=1$ and with a half-width of $\Delta = 2 H/r$. Throughout all our simulations, both N-body and hydro, we assumed a disk scale-height equal to $H/r=0.05$ and a depth of the cavity equal to 
$F=\Sigma_o/\Sigma_i = 10$. The density outside the cavity will be set to $\Sigma_o=N \Sigma_0$ with $N$ a positive integer and $\Sigma_0=10^{-3} M_{\odot} \ AU^{-2}$. In other words, we will perform simulations with different surface densities given as multiples of an arbitrary base value $\Sigma_0$. 

All plots shows the time evolution of the ratio of mean-motions $n_1/n_2$ between both planets. In the top plot we considered a fixed value of $\Sigma_o=15 \Sigma_0$, but varied the mass ratio of the bodies in the interval $m_2/m_1 \in [10^{-2},10^0]$. In all cases, the mass of the largest body was equal to one Saturn mass. 

Since $m_1$ has the smallest initial semimajor axis, as well as the largest mass, its orbit suffers a faster orbital decay, and is the first planet to reach the cavity edge. In all the runs, this body was trapped near the center of the density jump ($r \sim 1.05 r_0$) with a small eccentricity. Meanwhile, the outer approaches the cavity edge, encountering successive and increasingly strong mean-motion resonances in its path. The orbital evolution is then dictated by two opposing forces: (i) the differential disk torque that subtracts angular momentum and causes further orbital decay, and (ii) resonant perturbations with the inner mass that can generates conditions for a net increase in the angular momentum.  

The top plot of Figure \ref{fig1_p} shows that the system is stopped at different mean-motion resonances depending on the mass ratio. For small values of $m_2/m_1$ the two planets are trapped in the 3/1 MMR, while the 3/2 commensurability is favored for larger mass ratios. This is expected, since a larger outer mass requires stronger resonance perturbations to counteract the disk torque. For $m_2/m_1 \ge 0.7$, however, it appears no commensurability is sufficiently strong, and both bodies reach the cavity edge and evolve towards a stable coorbital configuration. 

From these results it then appears that coorbital planets may be formed at a density jump in the disk if the mass ratio is sufficiently close to unity. We checked that the same result was obtained for other mass values, considering planets ranging from Neptune to Jupiter analogues.

The middle plot of Figure \ref{fig1_p} now analyses the dependence on the value of the surface density $\Sigma_o$ outside the cavity, while maintaining the same cavity depth $F$. Results are shown for two equal mass planets ($m_1=m_2=m_{\rm Sat}$) and four different multiples of the base density $\Sigma_0$. We note that smaller densities favor resonance trapping in non-coorbital configurations (e.g. $5/4$ MMR), but for large values of $\Sigma_o$, typically above $\Sigma_1 \sim 10\Sigma_0$, coorbital systems are the usual outcome. Of course there is some dependence on the initial conditions, mainly on the initial separation of the bodies, but these results are surprisingly robust and are representative of most of our runs.

Finally, the bottom frame shows the dependence on the eccentricity damping force $F_{\rm damp}$. Although both the circularization time scale $\tau_e$ and the orbital decay time $\tau_a$  have the same dependence with the planetary mass and disk density (e.g. Tanaka and Ward 2004), it is possible to fictitiously modify the eccentricity damping by changing the tangential force suffered by the bodies. Our main aim in this experiment is to see how the dynamics may be affected by different values of $\tau_e/\tau_a$. The plot shows two simulations, one with the correct value of $\tau_e$ (black) and one in which the damping rate was reduced 
by a factor $10$ (blue). Although a coorbital configuration was achieved in both cases, the reduced damping generates a much larger amplitude of oscillation of the eccentricities which may ultimately lead to orbital instability. In fact, some of these simulations showed complex exchanges between different types of coorbital
solutions ($L_4$, $L_5$, etc).

\begin{figure}
\centerline{\includegraphics*[width=21pc]{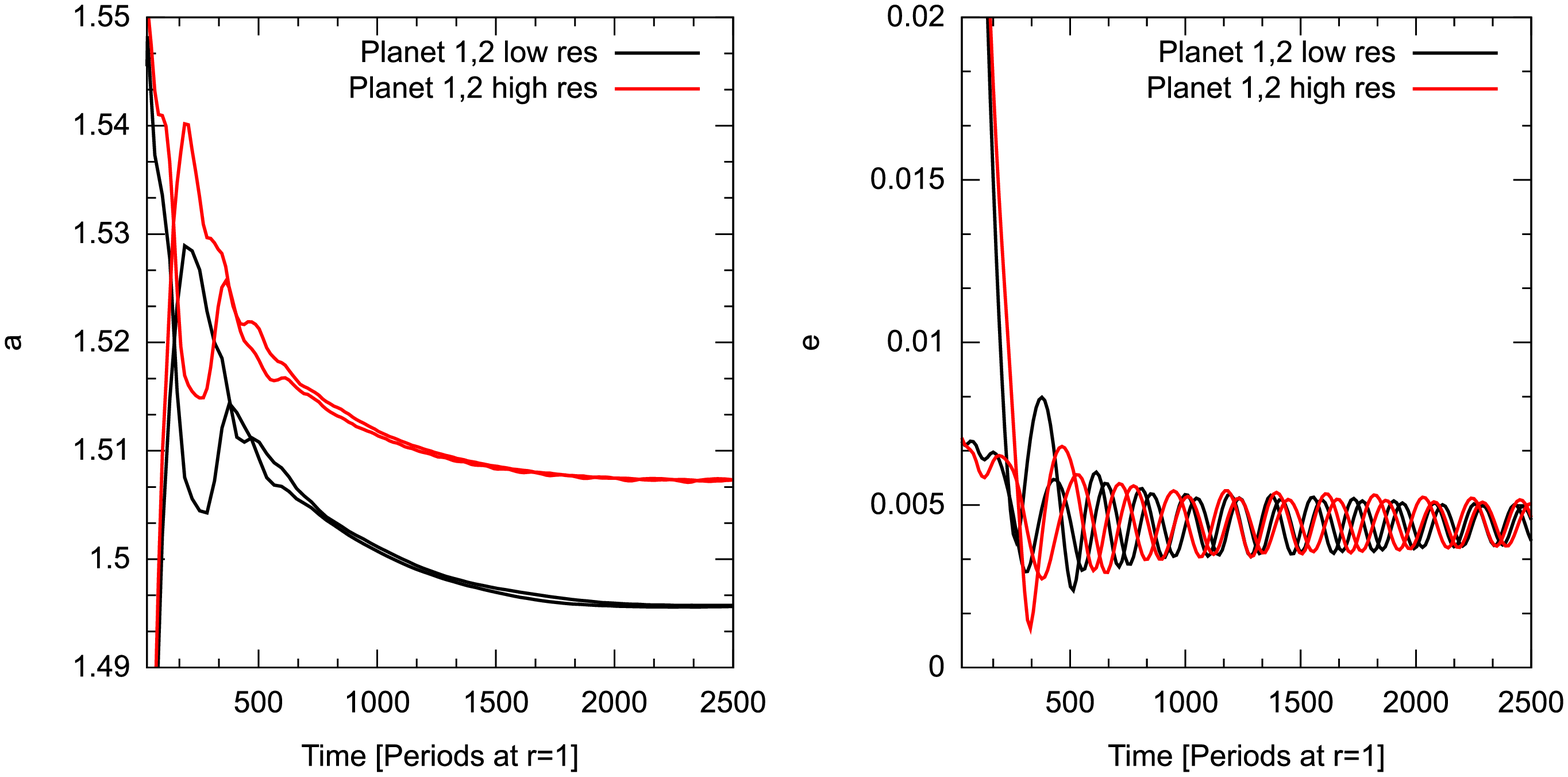}}
\caption{Comparison between two hydro-simulation with the same initial conditions (see text for details) but different grid resolutions. The high resolution simulation was constructed with 306 radial zones, while the low resolution run contained only 168 radial zones. Both lead to coorbital configurations with similar global dynamical features}
 \label{fig2_p}
\end{figure}

In conclusion, our preliminary N-body simulations appear to indicate that coorbital systems of massive planets can be obtained if the mass ratio is sufficiently close to unity and the surface density of the disk is sufficiently
high. Although the final outcomes have a certain dependence on the initial conditions (like all resonance trapping phenomena), we repeated the runs for initial separations between $0.5$ to $1.7 r_0$, finding no significant changes in the results. Inspired by these results, our next step is to upgrade to hydrodynamical simulations.

\subsection{Hydro-numerical setup}

\begin{figure}
\centerline{\includegraphics*[width=20pc]{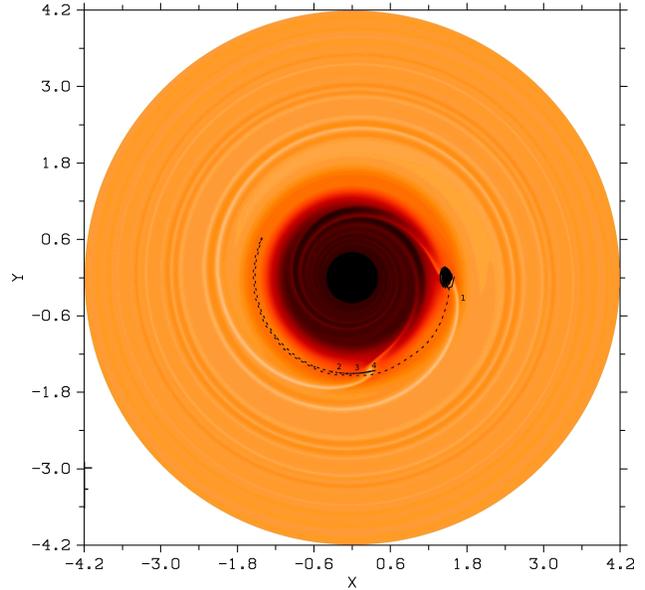}}
\caption{Snapshot of the final outcome of a high resolution (306 radial zones) hydro-simulation leading to two planets locked in a coorbital configuration. One of the planets ($m_1$) is fixed in the x-axis. The numbers 1, 2, 3 and 4 represent the location of the other planet ($m_2$) at different times of the run, with number 4 being the final configuration.}
 \label{fig4_p}
\end{figure}

Our hydro-simulations were carried out using the FARGO code (Masset, 2000) with a central star of one solar mass. We consider two-dimensional, locally isothermal disks uniformly distributed over a polar grid. 

The disk inner cavity (IC) was generated using an {\it ad hoc} step in kinematic viscosity around $r=r_0=1$, using the same recipe as described in Masset et al. (2006a). For all our runs we adopted a cavity depth of $F=\Sigma_{o}/\Sigma_{i}=10$, and a half-width of $\Delta = 0.4$. The kinematic viscosity was considered  uniform outside the cavity, and equal to $\nu=10^{-5}r_0^2\Omega_0^{-1}$, where $\Omega_0$ is the orbital frequency at radius $r_0$. The disk aspect ratio was chosen as $H/r=0.05$ and also constant over $r$. Boundary conditions were chosen to be non-reflecting for the inner edge of the disk. 

\begin{figure}
\centerline{\includegraphics*[width=15pc]{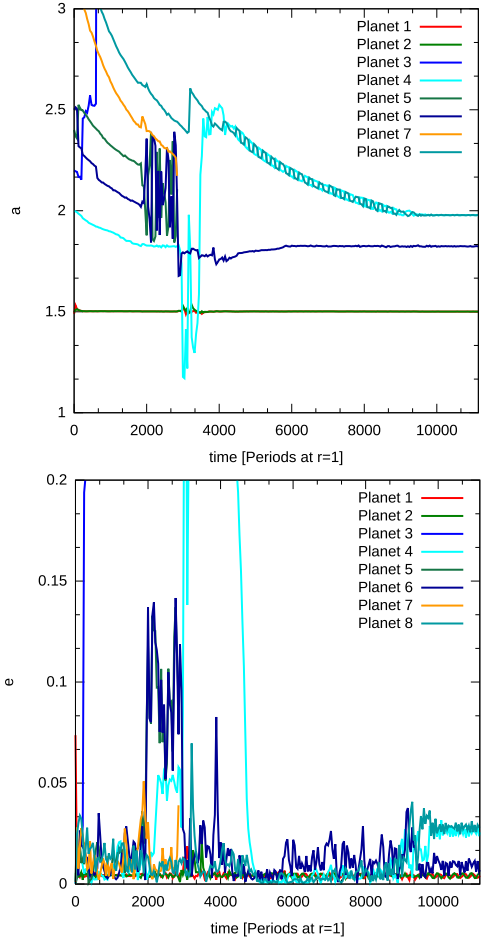}}
\caption{Results of a low resolution (168 radial zones) hydro-simulation of eight planets with masses in the Earth-Saturn range. Top plot shows the semimajor axis as function of time, while the lower frame shows the eccentricities.}
 \label{fig3_p}
\end{figure}

The disk was represented with a spatial resolution of 384 azimuthal zones and 168 radial zones. The inner radius of disk is set to $0.42r_0$ and the outer radius is set to $4.2r_0$. This resolution allows us to identify details as small as $0.04r_0$. Casoli \& Masset (2009) showed that the coorbital co-rotation torque exerted by the disk over a planet occurs in a scale of $x_s \sim 0.01 r_0$ for $m \sim 2 m_{\oplus}$, while Masset et al. (2006b) showed that the horse-shoe zone width in a two-dimentional disk scale as $x_s =1.16\sqrt{q/h}$. Consequently, $x_s$ should increase with the planetary mass. For example if we are interested in the dynamical evolution of Saturn-type planets, we expect that our choice for the spatial resolution should be adequate. 

To test this idea, we performed several simulations with the same initial conditions but different spatial resolutions for the disk, and compared the dynamical evolution of the planets. An example is shown in Figure \ref{fig2_p} for two Saturn-like planets initially in circular orbits far from the cavity edge and in non-resonant conditions. Both simulations show the same qualitative results, not only in the final outcome of the planets (both runs end in coorbital configurations) but also in the behavior of the eccentricities. The only noticeable change is a slight difference in the radial distance at which the planets are trapped. However, this is not significant and could be explained due to a decrease of the co-rotational torque from a low resolution description of the horse-shoe region.

\begin{figure}
\centerline{\includegraphics*[width=20pc]{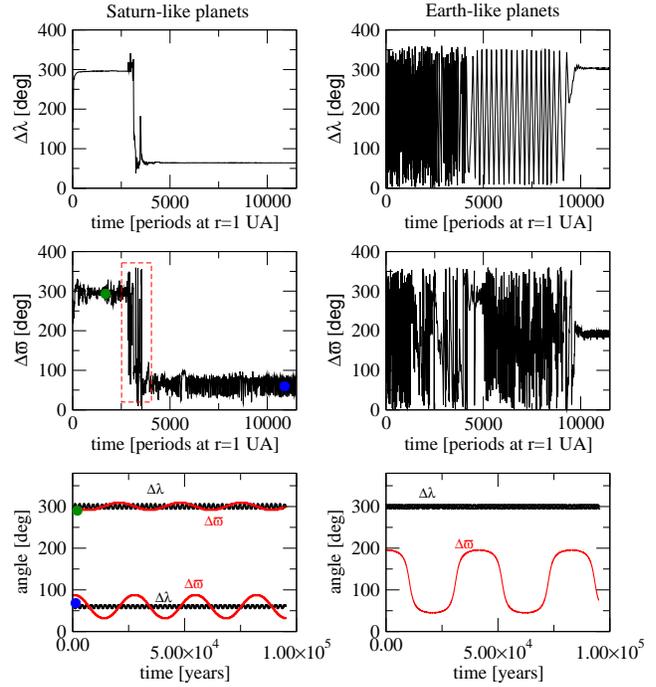}}
\caption{Evolution of characteristics angles for coorbital solutions (top frame $\Delta\lambda$ and middle frame $\Delta\varpi$ during the hydro-simulation. We reserve hand-left side to the system 1-2 and right side to the system 4-6. 
We plot in the lower frames the evolution of characteristics angles integrating the system without considering the interaction with the disk. At left frame two conditions were chosen, one at t=450 and the other at t=11450 (marked as color circles in the middle frame), meanwhile at right bottom frame the final configuration of earth-like planets was integrated.}
 \label{fig5_p}
\end{figure}

Following the predictions of our N-body runs, all our hydro-simulations were done with two equal-mass planets 
($m_1=m_2$) in a high-density disk. Initial conditions were varied both with respect to the initial separation between the planets and the initial eccentricities. We also considered different planetary masses between one Earth-mass and one Saturn-mass. 

Under a wide variety of initial conditions, we found that the orbital evolution of two Saturn-mass planets ultimately lead to the formation of stable coorbital configurations. An example is shown in Figure \ref{fig4_p}, where both bodies are trapped in $L_4/L_5$-type motion with a libration of both the resonant angle $\sigma = \lambda_2 - \lambda_1$ and the difference in longitudes of pericenter $\Delta \varpi = \varpi_2 - \varpi_1$.

To test the robustness and stability of the coorbital solutions to additional perturbations, we performed a new series of runs including additional six Earth-mass planets initially far from the cavity edge. The idea behind this set of experiments is to analyze both the orbital evolution of these new bodies, as well as their effects on the two coorbitals. Typical results are shown in Figure \ref{fig3_p}, where each color curve corresponds to a different planet (see inlaid color code). Planets 1 and 2 (i.e. $m_1$ and $m_2$, respectively) are the original Saturn-mass bodies in coorbital motion, while planets 3 through 8 are the new Earth-size masses. Since this type of simulation included smaller masses and a larger population, we employed a high-resolution description of the disk. 

We perceived no significant instability in the coorbitals, and their configuration remained unaffected by the new masses. These, however, suffered several cases of mutual scattering and close encounters that caused the ejection of three of the masses. One of the close encounters (between planets 4 and 6) led to a temporary trapping of planet 4 in the cavity edge but finally led to the formation of a new coorbital system with planet 8. The role of scattering between planets as a formation mechanism for coorbitals has already been pointed out by Cresswell \& Nelson (2006). The final outcome of this system consists of an inner coorbital system, a single Earth-mass planet trapped in an exterior mean-motion resonance and an additional coorbital pair in another resonance. Although this appears a very complex multiple resonant configuration, we found no indication of long-term instability. 

Figure \ref{fig5_p} shows the dynamics of both pairs of coorbitals: planets 1 and 2 are displayed on the left-hand plots, while the coorbital system composed of planets 4 and 8 are shown on the right. Top graphs present the temporal evolution of the resonant angle $\sigma = \Delta \lambda = \lambda_2 - \lambda_1$ while the middle plots show $\Delta \varpi = \varpi_2 - \varpi_1$. Once planet 4 is scattered into the cavity, the resonance relation between planets 1 and 2 is temporarily disrupted, resulting in a short-lived circulation of both angles. However, once the Earth-mass planet is sent back outside the cavity, the coorbital configuration of the inner planets is reestablished, although around $L_5$ instead of $L_4$. We did some follow-up integrations without hydrodynamical interaction choosing two intermediate stages as initial conditions. The bottom left-hand plot shows the results choosing as initial conditions the system configuration at $t=450$ and $t=11450$, respectively. The behavior of both critical angles show a small amplitude libration around the equilibrium solutions. Between these two solutions the system experiments temporarily $AL_4$ and $AL_5$ configurations.

The dynamical evolution of the small-mass pair is more complex, and a stable configuration is only reached near the end of the simulation, and corresponds to an $AL_5$ type orbit with large amplitude of oscillation of $\Delta \varpi$ or about $150^\circ$.

\section{Conclusions}\label{conclusions}

In this paper we have analyzed the detectability and possible formation mechanism of hypothetical massive planets in stable coorbital configurations. So far there are no known extrasolar planetary systems containing coorbital bodies, which may imply that either these configurations are extremely difficult to form or to detect
from RV surveys. 

We have studied the detectability of three different types of coorbital motion (QS, $L_4/L_5$ and $AL_4/AL_5$), trying to evaluate possible bias in detections and identify what kind of compatible configuration could be detected. The analysis of Keplerian contributions to radial velocities allowed us to predict the value for the signal of one single planet that could be confused with coorbital configurations.

For even low values of errors in radial velocities measurements ($\epsilon$ = 3 m/s) and observation time-spans covering four orbital periods (which include most of the presently detected exoplanets), coorbital configurations appear hard to identify: the results of the fitting process could easily confuse the RV data with that stemming from other configurations/systems (single planet or 2/1 MMR system). For large mass ratios, a correct identification of coorbital configuration is even more complicated and more easy to confuse with the other configurations. In all cases, the residuals of the different systems are comparable, even more so for large mass ratios (see Fig \ref{rms-stat}). 

Observation data sets covering longer time-spans allow to detect mutual perturbations, but once again the best fits are not always associated to coorbital motion. We have found several cases where other resonant configurations (i.e. 3/1 commensurability) actually give smaller residuals and better results. 

Coorbital motion is not only sensitive to the data set, but also to the fitting procedure. Nested two-planet strategies may also confuse the real dynamics and yield results widely different from the nominal orbits; in this sense simultaneous two-planet fits seem more robust. 

Transit observations should help distinguish coorbitals planets from other solutions.

Finally, we have found that stable coorbital systems with two massive planets may be formed from originally non-resonant orbits through their interaction with a inner cavity in the protoplanetary disk, as long as the surface density of the disk is sufficiently large. Both our N-body and hydro-simulations indicate a preference towards coorbitals with similar masses (i.e. $m_1 \sim m_2$). In all our simulations with large mass ratios, the smaller planet was either pushed inside the cavity or trapped in a mean-motion commensurability outside the density jump.

\section*{Acknowledgments}

This work has been supported by the Argentinian Research Council -CONICET- and the Funda\c{c}\~{a}o para a Ci\^{e}ncia e a Tecnologia (FCT) of Portugal.

\section*{Appendix. About the scale of simulations} \label{Appendix}

All our simulations were performed with a density jump centered at $r_0=1$. If we wish relate our results to anothor scale, for example $r_0=0.01$, we need to modify not only the spatial scale of the orbital system but also the surface density value. Here we present a simple recipe for the scale transformations in order to preserve the complete dynamics, including both the gravitational interactions and migration timescale. 

The equation of motion of a planet with mass $m_i$ interacting gravitationally with other $N$ bodies of mass $m_j$ is given by 
\begin{equation}
 \frac{d^2\mathbf{r_i}}{dt^2}=-\sum_{j\neq i}{\frac{Gm_j}{r_{ij}^2}}\mathbf{e_{ij}}
\end{equation}
where $\mathbf{e_{ij}}$ are the unit vectors corresponding to the relative positions. If we introduce a scale change in the  coordinates defined by $r=\alpha r'$ and a temporal transformation defined by $t=\beta t'$, then the gravitational dynamics is invariant if $\beta=\alpha^{3/2}$. 

In our problem, however, the dynamical evolution of the planets also includes their interaction with the gas disk, which is specified by the gradient of the total differential torque $\Gamma$. If we then desire to preserve the complete dynamics of the system under the spatial rescaling, then the torque must scale as:
\begin{equation}
\label{eq_a_1}
\Gamma = \alpha^{-1} \Gamma' .
\end{equation}

For a type I migration, the total torque is given by $\Gamma(r) \propto \Sigma(r) n^2 r^4$, where $n$ is the mean motion. We also assume that the surface density is given by a power law expression of type:
\begin{equation}
\Sigma(r)=\Sigma_o\left(\frac{r}{r_0}\right)^{\eta}
\end{equation}
for a suitable exponent $\eta$. The coefficient $\Sigma_o$ denotes the surface density at $r=r_0$. Then, if we apply the spatial and temporal transformations required to preserve the gravitational dynamics, we obtain
\begin{equation}
 \Gamma = \alpha \Sigma_o \left(\frac{r'}{r'_0}\right)^{\eta} n'^2 r'^4 
               = \alpha \left(\frac{\Sigma_o}{\Sigma'_o}\right) \Gamma' .
\end{equation}
Therefore, for scale our simulation to a new suitable value of $r_0$, it is necessary to do the transformation:
\begin{eqnarray}
 r&=&\alpha r' \\
 t&=&\alpha^{3/2} t	' \\
 \Sigma_o&=&\alpha^{-2} \Sigma_o'
\end{eqnarray}
For this new radii, time and density value, the dynamics of system is invariant.

\newpage

\end{document}